\documentclass[letterpaper,12pt,notoc]{JHEP3}

\usepackage{graphicx}
\usepackage{amsmath}
\usepackage{amssymb}

\preprint{ \hbox{}\hfill arXiv:1007.5438}

\title{Vacua and RG flows in $N=9$ three dimensional gauged supergravity}
\author{Auttakit Chatrabhuti$^{a, \, b}$ and Parinya Karndumri$^{c,\, d}$\\
$^a$Theoretical High-Energy Physics and Cosmology Group, Department
of Physics, Faculty of Science, Chulalongkorn University, Bangkok 10330, Thailand\\
$^b$Thailand Center of Excellence in Physics, CHE, Ministry of Education, Bangkok 10400, Thailand \\
$^c$INFN, Sezione di Trieste, Italy\\
$^d$International School for Advanced Studies (SISSA), via Bonomea
265, 34136 Trieste, Italy\\
E-mail: \email{auttakit@sc.chula.ac.th},
\email{karndumr$@$sissa.it}}

\abstract{We study some vacua of $N=9$ three dimensional gauged
supergravity. The theory contains sixteen scalar fields
parametrizing the exceptional coset space
$\frac{F_{4(-20)}}{SO(9)}$. Various supersymmetric and some
non-supersymmetric AdS$_3$ vacua are found in both compact and
non-compact gaugings with gauge groups $SO(p)\times SO(9-p)$ for
$p=0, 1, 2, 3, 4$, $G_{2(-14)}\times SL(2)$ and $Sp(1,2)\times
SU(2)$. We also study many RG flow solutions, both analytic and
numerical, interpolating between supersymmetric AdS$_3$ critical
points in this theory. All the flows considered here are driven by a
relevant operator of dimension $\Delta=\frac{3}{2}$. This operator
breaks conformal symmetry as well as supersymmetry and drives the
CFT in the UV to another CFT in the IR with lower supersymmetries.}
\begin{document}
\section{Introduction}
Three dimensional Chern-Simons gauged supergravity has a very rich
structure. The theory admits gauge groups of various types namely
compact, noncompact, non semisimple and complex gauge groups
\cite{nicolai1, nicolai2, nicolai3}. This stems from the fact that
there is no restriction on the number of gauge fields. The gauge
fields are introduced to the gauged theory by a Chern-Simons kinetic
term which results in their non propagating nature in the theory.
This peculiar feature comes from the duality between vectors and
scalars in three dimensions. All the bosonic propagating degrees of
freedom are carried by the scalars because pure supergravity in
three dimensions is also topological. \\ \indent Maximal gauged
supergravity in three dimensions has been constructed in
\cite{nicolai1, nicolai2, nicolai3}. The construction of the $N=8$
theory can be found in \cite{N8}. All extended three dimensional
gauged supergravities with $N\leq 16$ have been given in a unique
formulation in \cite{dewit}. This is a gauged version of the
ungauged theory constructed in \cite{dewit1}. Vacua of these
theories have been studied in some details e.g. see \cite{Fisch,
N16Vacua} for $N=16$ theory and \cite{gkn, bs} for $N=4$ and $N=8$
theories. \\ \indent In three dimensional supergravities with $N>4$,
the scalar target space manifold is a symmetric space and can be
written as a coset space $\frac{G}{H}$, where $G$ is a global
symmetry group, and $H$ is its maximal compact subgroup. For the
theories with $N>8$, the target space is unique because there is
only one supermultiplet in these cases \cite{dewit1}. In this paper,
we study $N=9$ gauged theory in which the scalar manifold is given
by the exceptional coset $\frac{F_{4(-20)}}{SO(9)}$. We will study
some vacua of this theory with gauge groups $SO(p)\times SO(9-p)$
for $p=0,1,2,3,4$, $G_{2(-14)}\times SL(2)$ and $Sp(1,2)\times
SU(2)$. All these gauge groups have been shown to be consistent
gaugings in \cite{dewit}. We will study some vacua of these gaugings
and give relevant superconformal groups for maximally supersymmetric
vacua with all scalars being zero.
\\ \indent The possibility to study holographic renormalization
group flows is one of the interesting consequences of the AdS/CFT
correspondence \cite{maldacena}. We will also study some
supersymmetric RG flow solutions interpolating between
supersymmetric AdS$_3$ vacua. In gauged supergravity, these
solutions are domain walls interpolating between critical points of
the scalar potential. They have an interpretation in the dual field
theory as an RG flow driving the UV CFT to the IR fixed point
corresponding to the CFT in the IR. In this paper, we will find flow
solutions in three dimensional gauged supergravity. The solutions
describe the RG flows in two dimensional field theories. Some
supersymmetric flow solutions in three dimensional gauged
supergravity have been studied in \cite{bs, gkn} for $N=8$ and $N=4$
theories, respectively. Some flow solutions of $N=2$ models have
been studied in \cite{deger}. In this paper, we give the analogous
analysis in the $N=9$ theory. The is the largest amount of
supersymmetry ever studied so far in the context of RG flows in
three dimensional gauged supergravities.
\\ \indent The paper is organized as follows. In section \ref{N9},
we review some useful ingredients to construct the $N=9$ gauged
theory. We also give the explicit construction of the gauged theory
with symmetric scalar manifold $\frac{F_{4(-20)}}{SO(9)}$ in detail.
The procedure can be applied to other theories with different values
of $N$ as well. Various vacua are found in section \ref{vacua}. We
then find some flow solutions in section \ref{flow}. Finally, we
give some conclusions and comments in section \ref{conclusion}.
\section{$N=9$ three dimensional gauged supergravity}\label{N9}
In this section, we construct $N=9$ three dimensional gauged
supergravity using the formulation given in \cite{dewit}. The $N=9$,
$SO(9)$ gauged three dimensional supergravity has also been
constructed in \cite{N9_nishino}, but we will follow the
construction of \cite{dewit} because this formulation can be easily
extended to other gauge groups. We start by reviewing
some formulae and all the ingredients needed in this paper. \\
\indent In symmetric spaces $G/H$, we have the following
decompositions of the $G$ generators $t^{\mathcal{M}}$ into
$\{X^{IJ}, X^\alpha, Y^A\}$. The maximal compact subgroup $H$ of
$G$, consists of the $SO(N)$ R-symmetry and an additional factor
$H'$ such that $H=SO(N)\times H'$. The scalar fields parametrizing
the target space are encoded in the coset representative $L$. This
transforms under global $G$ and local $H$ symmetries by
multiplications from the left and right, respectively. The latter
can be used to eliminate the spurious degrees of freedom such that
$L$ is parametrized by $\textrm{dim}\,(G/H)$ physical scalars. In
our case, the maximal compact subgroup of $G=F_{4(-20)}$ is $SO(9)$,
so there is no factor $H'$. Generators $X^{IJ}$, $I,J=1,2,\ldots N$
generate $SO(N)$, and $Y^A$, transforming in a spinor representation
of $SO(N)$, are non-compact generators of $G$. The target space has
a metric $g_{ij}$, $i,j=1,2,\ldots d=\textrm{dim}\, (G/H)$ given by
\begin{equation}
g_{ij}=e^A_ie^B_j \delta_{AB}\,.
\end{equation}
Extended supersymmetries are described by $N-1$ almost complex
structures $f^{Pi}_{\phantom{as}j}$, $P=2,\ldots, N$. We can
construct $SO(N)$ generators from these $f^{Pi}_{\phantom{as}j}$'s
by forming tensors $f^{IJ}_{ij}$ via \cite{dewit}
\begin{equation}
f^{PQ}=f^{[P}f^{Q]},\qquad f^{1P}=-f^{P1}=f^P\, . \label{almost
complex structures}
\end{equation}
The tensors $f^{IJ}_{ij}$, being generators of $SO(N)$ in the spinor
representation, are given in terms of $SO(N)$ gamma matrices by
\begin{equation}
f^{IJ}_{ij}=-\Gamma^{IJ}_{AB}e^A_ie^B_j\,.
\end{equation}
The indices $A$ and $B$ are tangent space or ``flat'' indices on the
scalar target space. The vielbein of the target space is encoded in
the expansion
\begin{equation}
L^{-1} \partial_i L= \frac{1}{2}Q^{IJ}_i X^{IJ}+Q^\alpha_i
X^{\alpha}+e^A_i Y^A\, .\label{cosetFormula1}
\end{equation}
$Q^{IJ}_i$ and $Q^\alpha_i$ are composite connections for $SO(N)$
and $H'$, respectively. \\ \indent The gaugings are described by the
gauge invariant embedding tensor $\Theta_{\mathcal{M}\mathcal{N}}$.
From $\Theta_{\mathcal{M}\mathcal{N}}$, we can compute the $A_1$ and
$A_2$ tensors as well as the scalar potential via the so-called
T-tensor using
\begin{eqnarray}
A_1^{IJ}&=&-\frac{4}{N-2}T^{IM,JM}+\frac{2}{(N-1)(N-2)}\delta^{IJ}T^{MN,MN},\nonumber\\
A_{2j}^{IJ}&=&\frac{2}{N}T^{IJ}_{\phantom{as}j}+\frac{4}{N(N-2)}f^{M(I
m}_{\phantom{as}j}T^{J)M}_{\phantom{as}m}+\frac{2}{N(N-1)(N-2)}\delta^{IJ}f^{KL\phantom{a}m}_{\phantom{as}j}T^{KL}_{\phantom{as}m},
\nonumber \\
V&=&-\frac{4}{N}g^2(A_1^{IJ}A_1^{IJ}-\frac{1}{2}Ng^{ij}A_{2i}^{IJ}A_{2j}^{IJ}).
\end{eqnarray}
The T-tensors are defined by
\begin{equation}
T_{\mathcal{A}\mathcal{B}}=\mathcal{V}^{\mathcal{M}}_{\phantom{a}\mathcal{A}}\Theta_{\mathcal{M}\mathcal{N}}
\mathcal{V}^{\mathcal{N}}_{\phantom{a}\mathcal{B}}\, .
\end{equation}
All the $\mathcal{V}$'s are given by
\begin{equation}
L^{-1}t^\mathcal{M}L=\frac{1}{2}\mathcal{V}^{\mathcal{M}IJ}X^{IJ}+\mathcal{V}^\mathcal{M}_{\phantom{a}\alpha}X^\alpha+
\mathcal{V}^\mathcal{M}_{\phantom{a}A}Y^A.\label{cosetFormula}
\end{equation}
Using these, we can now construct the $N=9$ theory. We give the
procedure in detail but leave some formulae to the appendix. We
begin with the $\frac{F_{4(-20)}}{SO(9)}$ coset. The 52 generators
of the compact $F_4$ have been explicitly constructed by realizing
$F_4$ as an automorphism group of the Jordan algebra $J_3$ in
\cite{F4}. There are 16 non-compact and 36 compact generators in
$F_{4(-20)}$. Under $SO(9)$, the 52 generators decompose as
\begin{displaymath}
\mathbf{52}\rightarrow \mathbf{36}+\mathbf{16}
\end{displaymath}
where $\mathbf{36}$ and $\mathbf{16}$ are adjoint and spinor
representations of $SO(9)$, respectively. The non-compact
$F_{4(-20)}$ can be obtained from the compact $F_4$ by using ``Weyl
unitarity trick'', see \cite{exceptional coset} for an example with
$G_2$. This is achieved by introducing a factor of $i$ to each
generator corresponding to the non-compact generators. From
\cite{F4}, the compact subgroup $SO(9)$ is generated by, in the
notation of \cite{F4}, $c_1,\ldots, c_{21}$, $c_{30},\ldots,
c_{36}$, $c_{45},\ldots, c_{52}$. We have chosen the same $SO(9)$
subgroup as in \cite{F4} among the three possibilities, see
\cite{F4} for a discussion. The remaining 16 generators are our
non-compact ones which we will define by
\begin{equation}
Y^A=\Bigg\{
\begin{array}{rl}
ic_{A+21} &\textrm{for} \; A=1, \ldots, 8\\
ic_{A+28} &\textrm{for} \; A=9, \ldots, 16
\end{array} .
\end{equation}
Note that the $SO(9)$ generators $c_i$ in \cite{F4}, are labeled by
the $F_4$ adjoint index.  In order to apply the $SO(9)$ covariant
formulation of $N=9$ theory, we need to relabel them by using the
$SO(9)$ antisymmetric tensor indices i.e. $X^{IJ}$.  To do this, we
first note the relevant algebra from \cite{dewit}
\begin{equation}
[t^{IJ},t^{KL}]=-4\delta^{[I[K}t^{L]J]}, \qquad
[t^{IJ},t^A]=-\frac{1}{2}f^{IJ,AB}t_B, \qquad [t^A,t^B]=
\frac{1}{4}f^{AB}_{IJ}t^{IJ}\label{algebra}
\end{equation}
where we have used the flat target space indices in $f^{IJ}_{AB}$
and the non-compact generators, $t^A$. Using the first commutator in
\eqref{algebra}, we can map all $c_i$'s forming $SO(9)$ to the
desired form $X^{IJ}$. The detail of this is given in the appendix.
The next step is to find the $f^{IJ}$. In order to be compatible
with the $F_4$ algebra given in \cite{F4}, we need to use the second
and the third commutators in \eqref{algebra} to extract the
component of $f^{IJ}_{AB}$ rather than putting the explicit forms of
gamma matrices from another basis. There are eight independent
$f^{IJ}$ from which all other components follow from \eqref{almost
complex structures}. We will not give all of the $f^{IJ}$ here due
to their complicated form. \\ \indent We now come to various
gaugings characterized by the embedding tensors $\Theta$. The
embedding tensors for the compact gaugings with gauge groups
$SO(p)\times SO(9-p)$, $p=0,\ldots , 4$ are given by \cite{dewit}
\begin{equation}
\Theta_{IJ,KL}=\theta \delta^{KL}_{IJ}+\delta_{[I[K}\Xi_{L]J]}
\end{equation}
where
\begin{equation}
\Xi_{IJ}=\Bigg\{
\begin{array}{rl}
2\big(1-\frac{p}{9}\big)\delta_{IJ} & \textrm{for} \; I \leq p\\
-2\frac{p}{9}\delta_{IJ} & \textrm{for} \; I > p
\end{array} , \qquad \theta=\frac{2p-9}{9}\, .
\end{equation}
There is only one independent coupling constant, $g$. The gauge
generators can be easily obtained from $SO(9)$ generators $X^{IJ}$
by choosing appropriate values for the indices $I,J$. For example,
in the case of $SO(2)\times SO(7)$ gauging, we have the following
gauge generators
\begin{eqnarray}
SO(7)&:& T_1^{ab}=X^{ab},\,\,\, a,b=1,\ldots 7, \nonumber \\
SO(2)&:& T_2=X^{89}\, .
\end{eqnarray}
\indent We then move to non-compact gaugings with gauge groups
$G_{2(-14)}\times SL(2)$ and $Sp(1,2)\times SU(2)$. We find the
following embedding tensors
\begin{eqnarray}
G_{2(-14)}\times SL(2)&:&
\Theta_{\mathcal{M}\mathcal{N}}=\eta^{G_2}_{\mathcal{M}\mathcal{N}}-\frac{1}{6}\eta^{SL(2)}_{\mathcal{M}\mathcal{N}},\label{ThetaG2}\\
Sp(1,2)\times
SU(2)&:&\Theta_{\mathcal{M}\mathcal{N}}=\eta^{Sp(1,2)}_{\mathcal{M}\mathcal{N}}-12\eta^{SU(2)}_{\mathcal{M}\mathcal{N}}\label{ThetaSp}
\end{eqnarray}
where $\eta^{G_0}$ is the Cartan Killing form of the gauge group
$G_0$. The gauge generators in these two gaugings are given in the
appendix. \\ \indent Using these embedding tensors and equation
\eqref{cosetFormula}, we can find all the $\mathcal{V}$'s and
T-tensors. With the help of the computer algebra system
\textsl{Mathematica} \cite{Mathematica}, it is then straightforward
to compute $A_1$ and $A_2$ tensors and finally the scalar potential
for each gauge group. In the next section, we will give all of these
potentials but refer the readers to the appendix for $\mathcal{V}$'s
and T-tensors. For completeness, we also give here the condition for
finding stationary points of the potential. We are most interested
in supersymmetric AdS$_3$ vacua, so we mainly work with the
condition for supersymmetric stationary points. As given in
\cite{dewit}, see also \cite{nicolai2} for $N=16$, the
supersymmetric stationary points satisfy the two equivalent
conditions
\begin{eqnarray}
A_{2i}^{JI}\epsilon^J&=&0\nonumber\\
\textrm{and}\qquad A_1^{IK}A_1^{KJ}\epsilon^J&=&
-\frac{V_0}{4g^2}\epsilon^I=\frac{1}{N}(A_1^{IJ}A_1^{IJ}-\frac{1}{2}Ng^{ij}A_{2i}^{IJ}A_{2i}^{IJ})\epsilon^I,\label{a1a2condition}
\end{eqnarray}
where $V_0$ is the value of the potential at the critical point i.e.
the cosmological constant. $\epsilon^I$ are the Killing spinors
corresponding to the residual supersymmetries at the stationary
point. The second condition simply says that $\epsilon^I$ is an
eigenvector of $A_1$ with an eigenvalue $\sqrt{\frac{-V_0}{4g^2}}$
or $-\sqrt{\frac{-V_0}{4g^2}}$. In addition, these two conditions
are indeed equivalent as shown in \cite{dewit}.
  \\ \indent The condition for any stationary points, not
necessarily supersymmetric, is \cite{dewit}
\begin{equation}
3A_1^{IK}A^{KJ}_{2j}+Ng^{kl}A^{IK}_{2k}A^{KJ}_{3lj}=0\label{extremum}
\end{equation}
where $A^{KL}_{3lj}$ is defined by
\begin{equation}
A^{IJ}_{3ij}=\frac{1}{N^2}\bigg[-2D_{(i}D_{j)}A^{IJ}_1+g_{ij}A^{IJ}_1+A^{K[I}_1f^{J]K}_{ij}+2T_{ij}\delta^{IJ}-4D_{[i}T^{IJ}_{\phantom{as}
j]}-2T_{k[i}f^{IJk}_{\phantom{asd}j]}\bigg].
\end{equation}
For supersymmetric critical points, we will mostly work with the two
equivalent conditions given by \eqref{a1a2condition}. However, for
non-supersymmetric points, the condition \eqref{extremum} is
necessary to ensure that all the points are indeed stationary
points.
\section{Vacua of $N=9$ gauged supergravity}\label{vacua}
In this section, we give some vacua of the $N=9$ gauged theory with
the gaugings mentioned in the previous section. We will discuss the
isometry groups of the background with maximal supersymmetries at
$L=\mathbf{I}$. This is a supersymmetric extension of the
$SO(2,2)\sim SO(1,2)\times SO(1,2)$ isometry group of AdS$_3$. The
superconformal group can be identified by finding its bosonic
subgroup and representations of supercharges under this group. A
similar study has been done in \cite{N16Vacua} for models with
$N=16$ supersymmetry. The full list of superconformal groups in two
dimensions can be found in \cite{SC}. We first start with compact
gaugings.
\subsection{Vacua of compact gaugings}
It has been shown in \cite{warner} that the critical points obtained
from the potential restricted on a scalar manifold which is
invariant under some subgroups of the gauge group are critical
points of the full potential. This invariant manifold is
parametrized by all scalars which are singlets under the chosen
symmetry. To make things more manageable, we will not study the
scalar potential with more than four scalars. We choose to
parametrize the scalars by using the coset representative
\begin{equation}
L=e^{a_1Y_1}e^{a_2Y_2}e^{a_3Y_{15}}e^{a_4Y_{16}}\,
.\label{compact_coset}
\end{equation}
For any invariant manifold with the certain residual symmetry, our
choice for $L$ in \eqref{compact_coset} certainly does not cover the
whole invariant manifold. Therefore, the critical points on this
submanifold may not be critical points of the potential on the whole
scalar manifold. Nevertheless, we can use the argument of
\cite{warner} as a guideline to find critical points. After
identifying the critical points, we then use the stationarity
condition \eqref{extremum} to check whether our critical points are
truly critical points of the scalar potential.
\\ \indent
Let us identify some residual symmetries of \eqref{compact_coset}.
In $SO(9)$ gauging, with only $a_1\neq0$, $L$ has $SO(7)$ symmetry.
For $a_1,a_2\neq 0$, $L$ preserves $SO(6)$ symmetry. With
$a_1,a_2,a_3\neq0$ and $a_1,a_2,a_3,a_4\neq0$, $L$ preserves $SU(3)$
and $SU(2)$, respectively. In other gauge groups, $L$ will have
different residual symmetry. We will discuss the residual gauge
symmetry of each critical point, separately. We find that in all
cases, non trivial supersymmetric critical points arise with at most
two non zero scalars. With all four scalar fields turned on, the
conditions $A^{JI}_{2i}\epsilon^J=0$ are satisfied if and only if
two of the scalars vanish. So, we give below only potentials with
two scalars.
\\ \indent
In \eqref{compact_coset}, we have used the basis elements of $Y$'s
to parametrize each scalar field. We also find that, in this
parametrization, all the sixteen scalars are on equal footing in the
sense that any four of the $Y$'s among sixteen of them give the same
structure of the potential. As a consequence, any two non zero
scalars in \eqref{compact_coset} give rise to the same critical
points with the same location and cosmological constant. Notice that
this is not the case if we use different parametrization of $L$. For
example, by using linear combinations of $Y_i$'s as basis for the
four scalars in \eqref{compact_coset}, different choices of $Y_i$'s
in each basis may give rise to different structures of the scalar
potential.
\\ \indent
We use the same notation as in \cite{gkn} namely $V_0$ is the
cosmological constant, and $(n_-,n_+)$ refers to the number of
supersymmetries in the dual two dimensional filed theory. On the
other hand, the $n_+$ ($n_-$) corresponds to the number of positive
(negative) eigenvalues of $A_1^{IJ}$. For definiteness, we will keep
$a_1$ and $a_2$ non zero. Furthermore, we give the values of scalar
fields up to a trivial sign change.
\begin{itemize}
  \item $SO(9)$ gauging: \\
  The scalar potential is
  \begin{eqnarray}
  V&=&\frac{1}{32} g^2 (-1390-232 \cosh(2 a_1)+6 \cosh(4 a_1)+
  4 \cosh[2 (a_1-2 a_2)]\nonumber \\& &+4 \cosh(4 a_1-2 a_2)
   -112 \cosh[2 (a_1-a_2)]+\cosh[4 (a_1-a_2)]\nonumber \\& &
  -232 \cosh(2 a_2)+6 \cosh(4 a_2)-112 \cosh[2 (a_1+a_2)]
  \nonumber \\ & &+\cosh[4 (a_1+a_2)]+4 \cosh[2 (2 a_1+a_2)]
  +4 \cosh[2 (a_1+2 a_2)]).
  \end{eqnarray}
  This is the case in which the full R-symmetry group $SO(9)$ is gauged.
  There is no non trivial critical point with two scalars. For
  $a_2=0$, there are two critical points, but only the $L=\mathbf{I}$ solution has any supersymmetry.\\

  \begin{tabular}{|c|c|c|c|}
    \hline
    Critical points & $a_1$ & $V_0$ & Preserved supersymmetry \\ \hline
    1 & 0 & $-64g^2$ & (9,0) \\ \hline
    2 & $\cosh^{-1}2$ & $-100g^2$ & - \\
    \hline
  \end{tabular}\\

  The corresponding $A_1$ tensor at the supersymmetric point is
  \begin{equation}
  A_1^{(1)}=\textrm{diag}(-4,-4,-4,-4,-4,-4,-4,-4,-4)\, .
  \end{equation}
  The notation $A_1^{(1)}$ means that this is the value of the $A_1$ tensor
  evaluated at the critical point number 1 in the table.
  For $L=\mathbf{I}$, the background isometry is given by $Osp(9|2,\mathbb{R})\times
  SO(1,2)$. The non-supersymmetric critical point has unbroken $SO(7)$ gauge symmetry. This point is closely related
  to the non-supersymmetric $SO(7)\times SO(7)$ point found in
  $N=16$ $SO(8)\times SO(8)$ gauged supergravity \cite{N16Vacua}. Both
  the location and the value of the cosmological constants compared
  to the $L=\mathbf{I}$ point are very similar to that in
  \cite{N16Vacua}.

  \item $SO(8)$ gauging:\\
  The potential is
  \begin{eqnarray}
  V&=&-\frac{1}{16} g^2 [(26+2 \cosh(2 a_1)+\cosh[2 (a_1-a_2)]
  +2 \cosh(2 a_2)\nonumber \\ & &+\cosh[2
  (a_1+a_2)])^2
  -32 (\cosh^2a_2 \sinh^2(2 a_1)\nonumber \\ & &+\cosh^4a_1
  \sinh^2(2 a_2))].
  \end{eqnarray}
  This case is very similar to the $SO(9)$ gauging. There are two
  critical points with a single scalar.\\

  \begin{tabular}{|c|c|c|c|}
    \hline
    Critical points & $a_1$ & $V_0$ & Preserved supersymmetry \\ \hline
    1 & 0 & $-64g^2$ & (8,1) \\ \hline
    2 & $\cosh^{-1}2$ & $-100g^2$ & - \\
    \hline
  \end{tabular}
  \\

  The $A_1$ tensor is
  \begin{equation}
  A_1^{(1)}=\textrm{diag}(-4,-4,-4,-4,-4,-4,-4,-4,4)\, .
  \end{equation}
  For $L=\mathbf{I}$, the background isometry is given by $Osp(8|2,\mathbb{R})\times
  Osp(1|2,\mathbb{R})$. The critical point 2 is invariant under $G_2$ subgroup of $SO(8)$.
  Apart from the splitting of supercharges and residual gauge symmetry,
  the critical points in this gauging are the same as the $SO(9)$
  gauging.

  \item $SO(7)\times SO(2)$ gauging:\\ In this gauging, the
  potential is
  \begin{eqnarray}
  V&=&-\frac{1}{36864}g^2 [9 (342+40 \cosh a_1+18 \cosh(2 a_1)
  -4 \cosh(a_1-2 a_2)\nonumber \\ & &+16 \cosh(a_1-a_2)
  +3 \cosh[2 (a_1-a_2)]+12 \cosh(2 a_1-a_2)
  \nonumber \\ & &+8 \cosh a_2+50 \cosh(2 a_2)+16 \cosh(a_1+a_2)
  +3 \cosh[2 (a_1+a_2)]\nonumber \\ & &+12 \cosh(2 a_1+a_2)-4 \cosh(a_1
  +2 a_2))^2+8 (-576 \cosh^2\frac{a_2}{2} (-3\nonumber \\ & &+\cosh a_2
  -3 \cosh a_1 (1+\cosh a_2))^2 \sinh^2 a_1-9 (-1
  \nonumber \\ & &-8 \cosh a_1 (-1+\cosh a_2)+47 \cosh a_2+3 \cosh(2 a_1) (1+\cosh a_2)\nonumber \\ & &
  +6 \cosh^2 a_1 (1
  +\cosh a_2))^2 \sinh^2a_2)].
  \end{eqnarray}
  We find one supersymmetric critical point with
  \begin{equation}
  V_0=-144g^2, \qquad a_1=\cosh^{-1}\frac{5}{3},\qquad
  a_2=\cosh^{-1}2\,
  \end{equation}
  with the value of the $A_1$ tensor
  \begin{equation}
  A_1=\left(
\begin{array}{ccccccccc}
 -10 & 0 & 0 & 0 & 0 & 0 & 0 & 0 & 0 \\
 0 & -10 & 0 & 0 & 0 & 0 & 0 & 0 & 0 \\
 0 & 0 & -\frac{14}{3} & 0 & 0 & 0 & 0 & 0 & -\frac{8 \sqrt{2}}{3} \\
 0 & 0 & 0 & -10 & 0 & 0 & 0 & 0 & 0 \\
 0 & 0 & 0 & 0 & -10 & 0 & 0 & 0 & 0 \\
 0 & 0 & 0 & 0 & 0 & -10 & 0 & 0 & 0 \\
 0 & 0 & 0 & 0 & 0 & 0 & -10 & 0 & 0 \\
 0 & 0 & 0 & 0 & 0 & 0 & 0 & 6 & 0 \\
 0 & 0 & -\frac{8 \sqrt{2}}{3} & 0 & 0 & 0 & 0 & 0 & \frac{14}{3}
\end{array}
\right).
\end{equation}
  After diagonalization, we find
  \begin{equation}
  A_1=\textrm{diag}(-10, -10, -6, -10, -10, -10, -10, 6, 6).
  \end{equation}
  This is a (1,2) point with $SU(2)$ symmetry. With $a_2=0$, we find the following
  critical points\\

  \begin{tabular}{|c|c|c|c|}
    \hline
    Critical points & $a_1$ & $V_0$ & Preserved supersymmetry \\ \hline
    1 & 0 & $-64g^2$ & (7,2) \\ \hline
    2 & $\cosh^{-1}\frac{7}{3}$ & $-\frac{1024}{9}g^2$ & (0,1) \\
    \hline
  \end{tabular}.
  \\

   The corresponding values of the $A_1$ tensor are
   \begin{eqnarray}
   A_1^{(1)}&=&\textrm{diag}\left(-4,-4,-4,-4,-4,-4,-4,4,4\right) \nonumber \\
   \textrm{and}\,\,\, A_1^{(2)}&=&\textrm{diag}\left(-4,-4,-4,-4,-4,-4,-4,\frac{16}{3},8
            \right).
   \end{eqnarray}
   For $L=\mathbf{I}$, the background isometry is given by $Osp(7|2,\mathbb{R})\times
  Osp(2|2,\mathbb{R})$.
   The critical point 2 preserves $SU(3)$ symmetry. The location and
   value of the cosmological constant relative to the $L=\mathbf{I}$ point are similar to the $G_2\times
   G_2$ point in $SO(8)\times SO(8)$ gauged $N=16$ supergravity. In
   our result, the residual gauge symmetry is the $SU(3)$ subgroup
   of $G_2$ which is in turn a subgroup of $SO(7)$.

  \item $SO(6)\times SO(3)$ gauging:\\ We find the potential
  \begin{eqnarray}
  V&=&\frac{1}{128} g^2 (-3886-424 \cosh(2 a_1)+6 \cosh(4 a_1)+4 \cosh[2 (a_1
  -2 a_2)]\nonumber \\ & &+4 \cosh(4 a_1-2 a_2)-1536 \cosh(a_1-a_2)
  -208 \cosh[2 (a_1-a_2)]\nonumber \\ & &+\cosh[4 (a_1-a_2)]-424 \cosh(2
  a_2)
  +6 \cosh(4 a_2)\nonumber \\ & &-1536 \cosh(a_1+a_2)-208 \cosh[2 (a_1+a_2)]
  +\cosh[4 (a_1+a_2)]\nonumber \\ & &+4 \cosh[2 (2 a_1+a_2)]+4 \cosh[2 (a_1+2
  a_2)]).
  \end{eqnarray}
  One supersymmetric critical point is
  \begin{equation}
  V_0=-256g^2, \qquad a_1=\cosh^{-1}2,\qquad
  a_2=\cosh^{-1}3\,.
  \end{equation}
  with the value of the $A_1$ tensor
  \begin{equation}
  A_1=\left(
\begin{array}{ccccccccc}
 -16 & 0 & 0 & 0 & 0 & 0 & 0 & 0 & 0 \\
 0 & -16 & 0 & 0 & 0 & 0 & 0 & 0 & 0 \\
 0 & 0 & -4 & 0 & 0 & 0 & 0 & 0 & -4 \sqrt{3} \\
 0 & 0 & 0 & -16 & 0 & 0 & 0 & 0 & 0 \\
 0 & 0 & 0 & 0 & -16 & 0 & 0 & 0 & 0 \\
 0 & 0 & 0 & 0 & 0 & -16 & 0 & 0 & 0 \\
 0 & 0 & 0 & 0 & 0 & 0 & 8 & 0 & 0 \\
 0 & 0 & 0 & 0 & 0 & 0 & 0 & 8 & 0 \\
 0 & 0 & -4 \sqrt{3} & 0 & 0 & 0 & 0 & 0 & 4
\end{array}
\right).
\end{equation}
  This can be diagonalized to
  \begin{equation}
  A_1=\textrm{diag}\left(-16, -16, -8, -16, -16, -16, 8, 8,
  8\right).
  \end{equation}
  This is a (1,3) point and has $SO(3)\subset SO(6)$ symmetry.
  With $a_2=0$, we find the following
  critical points\\

  \begin{tabular}{|c|c|c|c|}
    \hline
    Critical points & $a_1$ & $V_0$ & Preserved supersymmetry \\ \hline
    1 & 0 & $-64g^2$ & (6,3) \\ \hline
    2 & $\cosh^{-1}3$ & $-144g^2$ & (0,2) \\
    \hline
  \end{tabular}.\\

  The corresponding values of the $A_1$ tensor are
   \begin{eqnarray}
   A_1^{(1)}&=&\textrm{diag}\left(-4,-4,-4,-4,-4,-4,4,4,4\right) \nonumber \\
   \textrm{and}\,\,\,
   A_1^{(2)}&=&\textrm{diag}\left(-10,-10,-10,-10,-10,-10,6,6,10\right).
   \end{eqnarray}
  For $L=\mathbf{I}$, the background isometry is given by $Osp(6|2,\mathbb{R})\times
  Osp(3|2,\mathbb{R})$. The critical point 2 is also invariant under
  $SO(3)$ subgroup of $SO(6)$.

  \item $SO(5)\times SO(4)$ gauging:\\ The potential for this
  gauging is
  \begin{eqnarray}
  V&=&\frac{1}{32} g^2 (3+\cosh a_1  \cosh a_2)^2 (-86+2 \cosh(2
  a_1)-24 \cosh(a_1-a_2)\nonumber \\ & &+\cosh[2 (a_1-a_2)]
  +2 \cosh(2 a_2)-24 \cosh(a_1+a_2)\nonumber \\ & &+\cosh[2
  (a_1+a_2)]).
  \end{eqnarray}
  There is no critical point with two non zero scalars.
  With $a_2=0$, we find the following
  critical points:\\

  \begin{tabular}{|c|c|c|c|}
    \hline
    Critical points & $a_1$ & $V_0$ & Preserved supersymmetry \\ \hline
    1 & 0 & $-64g^2$ & (5,4) \\ \hline
    2 & $\cosh^{-1}5$ & $-256g^2$ & (0,3) \\
    \hline
  \end{tabular}.\\

  The corresponding values of the $A_1$ tensor are
   \begin{eqnarray}
   A_1^{(1)}&=&\textrm{diag}\left(-4,-4,-4,-4,-4,4,4,4,4\right) \nonumber \\
   \textrm{and}\,\,\,
   A_1^{(2)}&=&\textrm{diag}\left(-16,-16,-16,-16,-16,8,8,8,16\right).
   \end{eqnarray}
  For $L=\mathbf{I}$, the background isometry is given by $Osp(5|2,\mathbb{R})\times
  Osp(4|2,\mathbb{R})$. The critical point 2 preserves
  $SO(4)_{\textrm{diag}}$ symmetry which is the diagonal subgroup of
  $SO(4)\times SO(4)$ with the first $SO(4)$ being a subgroup of
  $SO(5)$.
\end{itemize}
\subsection{Vacua of non-compact gaugings}
We now give some critical points of the non-compact gaugings. The
isometry group of the background with $L=\mathbf{I}$ consists of the
maximal compact subgroup of the gauge group and $SO(2,2)$ as the
bosonic subgroup. Using the generators given in the appendix, we can
compute the scalar potentials for these two gaugings. Notice that in
the non-compact gaugings, all sixteen scalars are not equivalent. At
the maximally symmetric vacua, the gauge group is broken down to its
maximal compact subgroup, and some of the scalars become Goldstone
bosons making some of the vector fields massive. This
``Higgs-mechanism'' results in the propagating $n_{ng}$ massive
vector fields where $n_{ng}$ denotes the number of non compact
generators which are broken at the critical point. The total number
of degrees of freedom remains the same because of the disappearance
of the $n_{ng}$ scalars, Goldstone bosons. For further detail, see
\cite{N16Vacua} in the context of $N=16$ models.
\begin{itemize}
  \item $G_{2(-14)}\times SL(2)$ gauging:\\
  The coset representative is chosen to be
  \begin{equation}
  L=e^{a_1Y_3}e^{a_2Y_{13}}\, .
  \end{equation}
  This parametrization has residual gauge symmetry $SU(2)$ which is
  a subgroup of $G_{2(-14)}$. With one of the scalars vanishing, $L$
  has $SU(3)$ symmetry.
  The potential with two scalars is given by
  \begin{eqnarray}
  V&=&\frac{1}{4608}g^2 [-23406-2520 \cosh(2 a_1)+70 \cosh(4 a_1)+
  8 \cosh(4 a_1-3 a_2)\nonumber \\ & &+28 \cosh[2 (a_1-2 a_2)]+
  28 \cosh(4 a_1-2 a_2)-560 \cosh[2 (a_1-a_2)]\nonumber \\ & &+
  \cosh[4 (a_1-a_2)]-1792 \cosh(2 a_1-a_2)+56 \cosh
  (4 a_1-a_2)\nonumber \\ & &+3472 \cosh(a_2)-6104 \cosh(2 a_2)-
  16 \cosh(3 a_2)+198 \cosh(4 a_2)\nonumber \\ & &-560 \cosh[2 (a_1+
  a_2)]+\cosh[4 (a_1+a_2)]-1792 \cosh(2 a_1+a_2)\nonumber \\ & &+
  28 \cosh[2 (2 a_1+a_2)]+56 \cosh(4 a_1+a_2)
  +28 \cosh[2 (a_1+2 a_2)]\nonumber \\ & &+8 \cosh(4 a_1+3 a_2)].
  \end{eqnarray}
  We find the following critical points:\\

  \begin{tabular}{|c|c|c|c|c|}
    \hline
    critical point & $a_1$ & $a_2$ & $V_0$ & preserved \\
    & & & & supersymmetries \\ \hline
    1 & 0 & 0 & $-\frac{64}{9}g^2$ & (7,2) \\ \hline
    2& 0 & $\cosh^{-1}\frac{1}{2}\sqrt{\frac{11+\sqrt{57}}{2}}$ & $-\frac{551+21\sqrt{57}}{72}g^2$ & - \\ \hline
    3 & $\cosh^{-1}2$ & 0 & $-\frac{100}{9}g^2$ & (0,1) \\ \hline
    4 & $\cosh^{-1}\frac{3}{2}$ & $\cosh^{-1}\frac{2}{\sqrt{3}}$ & $-\frac{1024}{81}g^2$ & (1,2) \\
    \hline
  \end{tabular}\\

  The corresponding values of the $A_1$ tensor are
   \begin{eqnarray}
   A_1^{(1)}&=&\textrm{diag}\left(-\frac{4}{3},-\frac{4}{3},-\frac{4}{3},-\frac{4}{3},-\frac{4}{3},
   -\frac{4}{3},-\frac{4}{3},\frac{4}{3},\frac{4}{3}\right), \nonumber \\
   A_1^{(3)}&=&\left(
\begin{array}{ccccccccc}
 -\frac{7 }{3} & 0 & 0 & 0 & 0 & 0 & 0 & 0 & 0 \\
 0 & -3 & 0 & 0 & 0 & 0 & \frac{2}{3} & 0 & 0 \\
 0 & 0 & -\frac{7}{3} & 0 & 0 & 0 & 0 & 0 & 0 \\
 0 & 0 & 0 & -\frac{7}{3} & 0 & 0 & 0 & 0 & 0 \\
 0 & 0 & 0 & 0 & -\frac{7}{3} & 0 & 0 & 0 & 0 \\
 0 & 0 & 0 & 0 & 0 & -\frac{7 }{3} & 0 & 0 & 0 \\
 0 & \frac{2}{3} & 0 & 0 & 0 & 0 & -3 & 0 & 0 \\
 0 & 0 & 0 & 0 & 0 & 0 & 0 & \frac{5}{3} & 0 \\
 0 & 0 & 0 & 0 & 0 & 0 & 0 & 0 & \frac{7}{3}
\end{array}
\right)
\end{eqnarray}
and
\begin{equation}
   A_1^{(4)}=\left(
\begin{array}{ccccccccc}
 -\frac{28 }{9} & 0 & 0 & 0 & 0 & -\frac{4 }{9} & 0 & 0 & 0 \\
 0 & -\frac{28 }{9} & 0 & 0 & 0 & 0 & \frac{4 }{9} & 0 & 0 \\
 0 & 0 & -\frac{8 }{3} & 0 & 0 & 0 & 0 & 0 & 0 \\
 0 & 0 & 0 & -\frac{14 }{9} & 0 & 0 & 0 & 0 & -\frac{2}{3} \sqrt{\frac{5}{3}}  \\
 0 & 0 & 0 & 0 & -\frac{8 }{3} & 0 & 0 & 0 & 0 \\
 -\frac{4 }{9} & 0 & 0 & 0 & 0 & -\frac{28 }{9} & 0 & 0 & 0 \\
 0 & \frac{4 }{9} & 0 & 0 & 0 & 0 & -\frac{28 }{9} & 0 & 0 \\
 0 & 0 & 0 & 0 & 0 & 0 & 0 & \frac{16}{9} & 0 \\
 0 & 0 & 0 & -\frac{2}{3} \sqrt{\frac{5}{3}} & 0 & 0 & 0 & 0 & \frac{14 }{9}
\end{array}
\right).
   \end{equation}
    $A_1^{(3)}$ and $A_1^{(4)}$ can be diagonalized to
    \begin{eqnarray}
    A_1^{(3)}&=&\textrm{diag}\left(-\frac{7}{3}, -\frac{11}{3}, -\frac{7}{3}, -\frac{7}{3}, -\frac{7}{3},
   -\frac{7}{3},-\frac{7}{3},\frac{5}{3},\frac{7}{3}\right),\nonumber \\
    A_1^{(4)}&=&\textrm{diag}\left(-\frac{32}{9}, -\frac{32}{9}, -\frac{8}{3}, -\frac{16}{9},
-\frac{8}{3},
   -\frac{8}{3}, -\frac{8}{3}, \frac{16}{9}, \frac{16}{9}\right).
   \end{eqnarray}
  For $L=\mathbf{I}$, the gauge group is broken down to its compact
  subgroup $G_2\times SO(2)$. The background isometry is given by $G(3)\times
  Osp(2|2,\mathbb{R})$. There are two $SU(3)$ points with completely broken supersymmetry (point 2)
  and (0,1) supersymmetry (point 3). Point 4 has $SU(2)$ symmetry.

  \item $Sp(1,2)\times SU(2)$ gauging:\\
  We choose the coset representative
  \begin{equation}
  L=e^{a_1(Y_1-Y_{10})}e^{a_2(Y_2+Y_9)}\, .
  \end{equation}
  This has symmetry $SO(3)\times SO(3)$ if any one of the scalars vanishes. This is the case in which our critical points lie. This
  symmetry is a subgroup of the $SO(5)\times SO(3)$ compact subgroup of $Sp(1,2)$ with the first $SO(3)$ being a subgroup of $SO(5)$.
  We find the potential
  \begin{eqnarray}
  V&=&\frac{1}{32} g^2 [-1390-232 \cosh(2 \sqrt{2} a_1)
  +6 \cosh(4 \sqrt{2} a_1)\nonumber \\ & &+4 \cosh[2 \sqrt{2} (a_1
  -2 a_2)]-112 \cosh[2 \sqrt{2} (a_1-a_2)]\nonumber \\ & &
  +\cosh[4 \sqrt{2} (a_1-a_2)]+4 \cosh[2 \sqrt{2} (2 a_1
  -a_2)]-232 \cosh(2 \sqrt{2} a_2)\nonumber \\ & &+6 \cosh(4 \sqrt{2} a_2)
  -112 \cosh
  [2 \sqrt{2} (a_1+a_2)]+\cosh[4 \sqrt{2} (a_1
  +a_2)]\nonumber \\ & &+4 \cosh[2 \sqrt{2} (2 a_1+a_2)]+4 \cosh[2
  \sqrt{2} (a_1+2 a_2)]].
  \end{eqnarray}
  Some of the critical points are given by\\

  \begin{tabular}{|c|c|c|c|c|}
    \hline
    critical point & $a_1$ & $a_2$ & $V_0$ & preserved supersymmetries \\ \hline
    1 & 0 & 0 & $-64g^2$ & (5,4) \\ \hline
    2& 0 & $\frac{\cosh^{-1}2}{\sqrt{2}}$ & $-100g^2$ & - \\ \hline
    3 & $\frac{\ln(2-\sqrt{3})}{\sqrt{2}}$ & 0 & $-100g^2$ & - \\ \hline
    4 & $\frac{\ln(2+\sqrt{3})}{\sqrt{2}}$ & 0 & $-100g^2$ & - \\
    \hline
  \end{tabular}\\

  with the corresponding $A_1$ tensor
  \begin{equation}
   A_1^{(1)}=\textrm{diag}\left(-4,-4,-4,-4,-4,4,4,4,4\right)
   \end{equation}
   for the critical point 1.
  For $L=\mathbf{I}$, the gauge group is broken down to its compact
  subgroup $Sp(1)\times Sp(2)\times SU(2)\sim SU(2)\times SO(5)\times
  SU(2)$. The two $SU(2)$'s factors combine to $SO(4)$ under which
  the right handed supercharges transform as $\mathbf{4}$. So, the background
  isometry is given by $Osp(5|2,\mathbb{R})\times
  Osp(4|2,\mathbb{R})$. Point 2, 3, and 4 are $SO(3)\times
  SO(3)$ points with completely broken supersymmetry.
\end{itemize}
We have checked that all critical points given above are truely
critical points of the corresponding potential. In the next section,
we will find RG flow solutions interpolating between some of these
vacua.
\section{RG flow solutions}\label{flow}
In this section, we study RG flow solutions in the $N=9$ theory
whose vacua are obtained in the previous section. We start by giving
the general formulae we will use in various gaugings. The strategy
to find supersymmetric flow solutions is to find the solutions to
the BPS equations coming from the supersymmetry transformations of
fermions which in this case, are $\delta\chi^{iI}$ and $\delta\psi^I_\mu$. \\
\indent We start by giving an ansatz for the metric
\begin{equation}
ds^2=e^{2A(r)}dx^2_{1,1}+dr^2\, .
\end{equation}
The relevant spin connection is
\begin{equation}
\omega_{\hat{\mu}}^{\phantom{a} \hat{\nu}\hat{r}}=A'\delta^\nu_\mu
\end{equation}
where hatted indices denote the tangent space indices, $\hat{\mu},
\hat{\nu}=0, 1$. We use the notation $'\equiv \frac{d}{dr}$ from now
on. We then recall the supersymmetry transformations from
\cite{dewit}
\begin{eqnarray}
\delta\psi^I_\mu
&=&\mathcal{D}_\mu\epsilon^I+gA_1^{IJ}\gamma_\mu\epsilon^J,\nonumber\\
\delta\chi^{iI}&=&
\frac{1}{2}(\delta^{IJ}\mathbf{1}-f^{IJ})^i_{\phantom{a}j}{\mathcal{D}{\!\!\!\!/}}\phi^j\epsilon^J
-gNA_2^{JIi}\epsilon^J\, .\label{susyvar}
\end{eqnarray}
We will not repeat the meaning of all the notations here but refer
the readers to \cite{dewit} for the detailed explanation. Using
\eqref{cosetFormula1}, we find, in our normalization,
\begin{equation}
\frac{d\phi^A}{dr}=\frac{1}{6}\textrm{Tr}(Y^AL^{-1}L').
\end{equation}
With this information, we are now in a position to set up the BPS
equations which are our flow equations. The $\delta\chi^{Ii}=0$
equation gives flow equations for the scalars while the
$\delta\psi^I_\mu=0$ is used to determine $A(r)$ in the metric. In
order to obtain the equation for $A(r)$, we impose
$\gamma_r\epsilon^I=\epsilon^I$, so the solution preserves half of
the original supersymmetries. We now apply this result to various
gaugings. In the gauging that admits a supersymmetric flow solution,
there must exist at least two AdS supersymmetric critical points
with different cosmological constants. The latter is related to the
central charge of the dual CFT as
\begin{equation}
c\sim \frac{1}{\sqrt{-V_0}}\, .
\end{equation}
According to the c-theorem, the c-function interpolating between the
central charges in the UV and IR fixed points is a monotonically
decreasing function along the flow from the UV to the IR. From the
previous section, there is no flow solution in the $SO(9)$, $SO(8)$
and $Sp(1,2)\times SU(2)$ gaugings because there is only one
supersymmetric critical point.
\subsection{RG flows in compact gaugings}
We start by finding flow solutions in the compact gaugings.
\subsubsection{$SO(7)\times SO(2)$ gauging}
\begin{figure}[!h] \centering
  \includegraphics[width=0.5\textwidth, bb = 0 0 200 150 ]{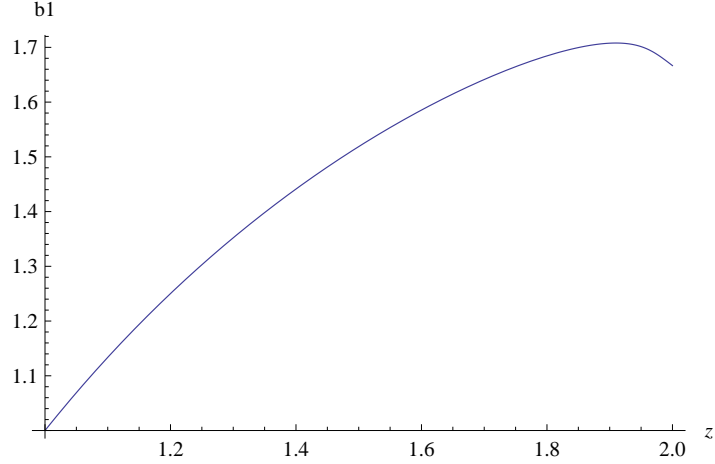}
  \caption{Solution for $b_1(z)$ in $SO(7)\times SO(2)$ gauging.}
  \label{Nsol1_27}
\end{figure}

\begin{figure}[!h] \centering
  \includegraphics[width=0.5\textwidth, bb = 0 0 200 150 ]{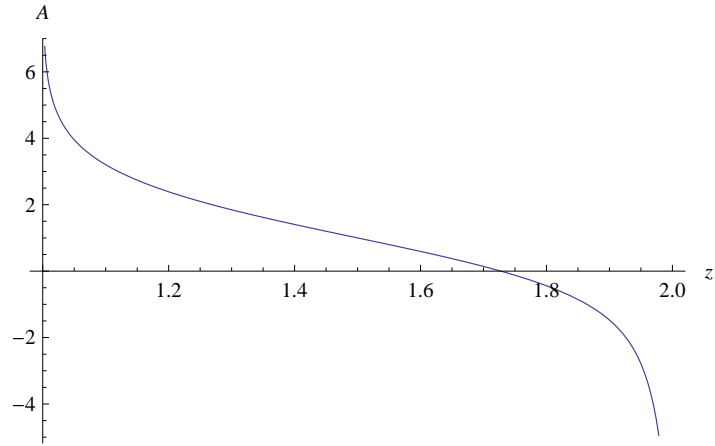}
  \caption{Solution for $A(z)$ in $SO(7)\times SO(2)$ gauging.}
  \label{Nsol2_27}
\end{figure}

  With a single scalar, the flow equation is given by
  \begin{equation}
  \frac{da_1}{dr}=g\sinh a_1(3\cosh a_1-7).
  \end{equation}
  Changing the variable to $b=\cosh a_1$, we find the solution
  \begin{equation}
  r=\frac{1}{20g}\ln (1+b)-\frac{1}{8g}\ln (b-1)+\frac{3}{40g}\ln
  (3b-7).\label{rsol}
  \end{equation}
  The supersymmetry transformation of the gravitino gives
  \begin{equation}
  \frac{dA}{dr}=-\frac{1}{2}g(b-5)(1+3b).
  \end{equation}
  We can solve this equation to obtain $A$ as a function of $b$
  using the equation for $\frac{db}{dr}$. The solution is
  \begin{equation}
  A=-\ln (b-1)-\frac{3}{10}\ln (1+b)+\frac{4}{5}\ln
  (3b-7).\label{Asol}
  \end{equation}
  In all these solutions, we have neglected all the additive
  constants to $A$ and $r$ because we can always shift $A$ and $r$
  to absorb them. From \eqref{rsol}, we see that as $a_1=0$, $r\rightarrow
  \infty$ and $r\rightarrow -\infty$ when $a_1=\cosh^{-1}\frac{7}{3}$. The UV
  point corresponds to $a_1=0$, and the IR point is at
  $a_1=\cosh^{-1}\frac{7}{3}$.
  The ratio of the central charges is given by
  \begin{equation}
  \frac{c_{\textrm{UV}}}{c_{\textrm{IR}}}=\sqrt{\frac{{V_0}_{\textrm{IR}}}{{V_0}_{\textrm{UV}}}}=\frac{4}{3}\,
  .
  \end{equation}
  At the UV point, the AdS$_3$ radius is $L=\frac{1}{8g}$. Near this point, the
  scalar fluctuation behaves as
  \begin{equation}
  \delta a_1\sim e^{-4gr}=e^{-\frac{r}{2L}}\, .
  \end{equation}
  Using the argument in \cite{muck, kw}, we find that the flow is driven by a relevant operator of dimension $\Delta=\frac{3}{2}$. In
  the IR, we find
  \begin{equation}
  \delta a_1\sim e^{\frac{5r}{4L}}, \qquad L=\frac{3}{32g}\, .
  \end{equation}
  The corresponding operator is irrelevant with dimension
  $\Delta=\frac{13}{4}$.
  The UV and IR points have supersymmetries (7,2)
  and (0,1), respectively. Our scalars are canonically normalized as
  can be easily checked by looking at the scalar kinetic terms, so
  we can directly read off the value of $m^2$ from the potential.
  Near the UV point, we find
  \begin{equation}
  V=-64g^2-24g^2a_1^2\, .
  \end{equation}
  The mass squared in unit of $\frac{1}{L^2}$ is
  $m^2L^2=-\frac{3}{4}$. The mass-dimension formula
  $\Delta(\Delta-2)=m^2L^2$ gives $\Delta=\frac{3}{2}$ in agreement
  with what we have found from the behavior of the scalar near the
  critical point. At the IR point, we find
  \begin{equation}
  V=-\frac{1024}{9}g^2+\frac{2080}{9}g^2a_1^2\, .
  \end{equation}
  The mass squared is $m^2L^2=\frac{65}{16}$ which gives precisely
  $\Delta=\frac{13}{4}$.
  \\ \indent We now consider a flow solution with two non zero scalars.
  Unfortunately, we are not able to find an analytic solution in
  this case. We do find a numerical solution interpolating between
  maximal supersymmetric point at $L=\mathbf{I}$ and the
  non trivial critical point with two scalars given in the previous
  section. We start by giving flow equations
  \begin{eqnarray}
  \frac{da_1}{dr}&=&g\frac{e^{\frac{a_2}{2}}\cosh{\frac{a_2}{2}}\sinh{a_1}}{1+e^{a_2}}[3\cosh{a_1}(1+\cosh{a_2})-\cosh{a_2}-13], \label{eqa1_27}\\
  \frac{da_2}{dr}&=&\frac{g}{16} (-65+9 \cosh^2a_1 (1+\cosh a_2)
  -8 \cosh a_1 (7+\cosh a_2)\nonumber \\ & &+3 \sinh^2 a_1
  +\cosh a_2 (47+3 \sinh^2a_1)) \sinh a_2\, . \label{eqa2_27}
  \end{eqnarray}
  Changing the variables to
  \begin{equation}
  a_1=\cosh^{-1}b_1,\qquad a_2=\cosh^{-1}b_2,
  \end{equation}
  we can rewrite \eqref{eqa1_27} and \eqref{eqa2_27} as
  \begin{eqnarray}
  b_1'&=&\frac{g}{2}(b_1^2-1)[3b_1(1+b_2)-b_2-13],\\
  b_2'&=&\frac{g}{4}(b_2^2-1)[11b_2-17+3b_1^2(1+b_2)-2b_1(7+b_2)].
  \end{eqnarray}
  It can be easily checked that $b_1=\frac{5}{3}, \, b_2=2$ is a
  fixed point of these equations. In order to find a numerical
  solution, we set $g=1$ and $b_2=z$. Taking $b_1$ as a function of
  $z$, we can write the two equations as a single equation
  \begin{equation}
  \frac{db_1}{dz}=\frac{2 (-13-z+3 (1+z) b_1) \left(-1+b_1^2\right)}{\left(-1
  +z^2\right) \left(-17+11 z-2 (7+z) b_1+3 (1+z) b_1^2\right)}\, .
  \end{equation}
  The numerical solution to this equation is shown in Figure
  \ref{Nsol1_27}. The gravitino variation gives
  \begin{equation}
  \frac{dA}{dr}=-\frac{1}{8} g [3-34 z+11 z^2-2 (13+14 z+z^2) b_1+3 (1+z)^2
  b_1^2]
  \end{equation}
  or
  \begin{equation}
  \frac{dA}{dz}=\frac{ (3-34 z+11 z^2-2 (13+14 z+z^2) b_1+
  3 (1+z)^2 b_1^2)}{2 (-1+z^2) (-17+11 z-2 (7+z)
  b_1+3 (1+z) b_1^2)}\, .
  \end{equation}
  The numerical solution for $A$ is shown in Figure \ref{Nsol2_27}.
  The UV point is at $r\rightarrow \infty$ and has (7,2)
  supersymmetries. The IR point has (1,2) supersymmetries and
  corresponds to $r\rightarrow -\infty$. The ratio of the central
  charges is
  \begin{equation}
  \frac{c_{\textrm{UV}}}{c_{\textrm{IR}}}=\frac{3}{2}\, .
  \end{equation}
  The behavior of the fluctuations of $a_1$ and $a_2$ near the fixed point can be found
  by linearizing \eqref{eqa1_27} and \eqref{eqa2_27}. We find
  \begin{equation}
  \delta a_1\sim e^{-\frac{r}{2L}},\qquad \delta a_2\sim
  e^{-\frac{r}{2L}},
  \end{equation}
  near the UV point with $L=\frac{1}{8g}$. We see that the flow is driven by a relevant operator of dimension $\frac{3}{2}$.
  Near the IR point with
  $L=\frac{1}{12g}$, we find
  \begin{equation}
  \delta a_1\sim \delta a_2 \sim e^{4gr}=e^{\frac{r}{3L}}\, .
  \end{equation}
  So, the operator becomes irrelevant at the IR and has dimension
  $\Delta=\frac{7}{3}$. We can also check this by computing the scalar
  masses from the potential although it is more complicated in this case because we will need to diagonalize
  the mass matrix. We only give the analysis at the UV point. The
  potential is fortunately diagonal and given by
  \begin{equation}
  V=-64g^2-24g^2(a_1^2+a_2^2)\, .
  \end{equation}
  We find $m^2L^2=-\frac{3}{4}$ which gives $\Delta=\frac{3}{2}$.
  \\ \indent In all other gaugings studied here, the same pattern
  appears, and the analysis is the same. So, we will quickly go through these
  cases and give only the main results without giving all the
  details. In particular, we will not give the scalar masses. These
  can be worked out as above.
\subsubsection{$SO(6)\times SO(3)$ gauging}
\begin{figure}[!h] \centering
  \includegraphics[width=0.5\textwidth, bb = 0 0 200 150 ]{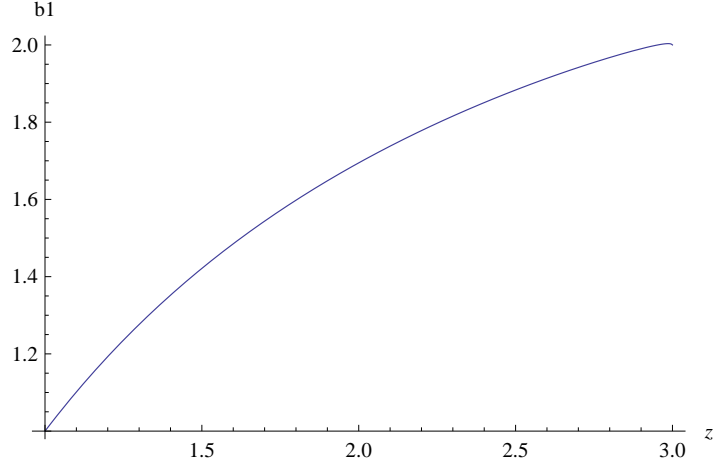}\\
  \caption{Solution for $b_1(z)$ in $SO(6)\times SO(3)$ gauging.}
  \label{sol1_36}
\end{figure}

\begin{figure}[!h] \centering
  \includegraphics[width=0.5\textwidth, bb = 0 0 200 150 ]{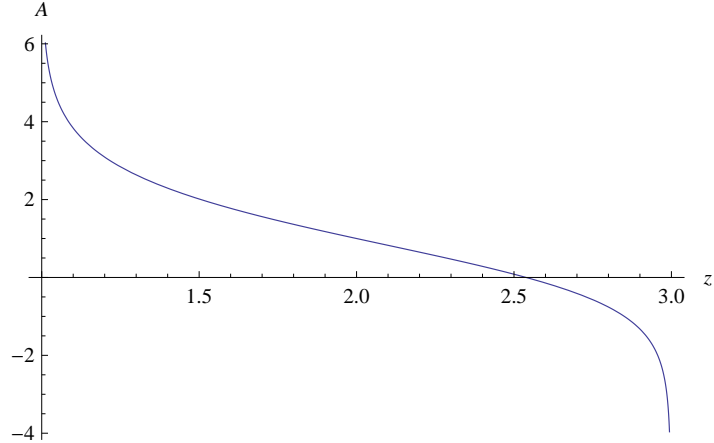}\\
  \caption{Solution for $A(z)$ in $SO(6)\times SO(3)$ gauging.}
  \label{sol2_36}
\end{figure}

We begin with a flow with one scalar. The flow equation is
\begin{equation}
a_1'=g(2\cosh{a_1}-6)\sinh{a_1}\, .
\end{equation}
With $a_1=\cosh^{-1}b$, we find
\begin{equation}
b'=2g(3-b-3b^2+b^3).
\end{equation}
This can be solved directly and gives
\begin{equation}
r=\frac{1}{16}\ln (9+6b-3b^2)-\frac{1}{8}\ln (b-1).
\end{equation}
The gravitino variation gives
\begin{equation}
A'=-\frac{g}{2}(\cosh{(2a_1)}-12\cosh a_1-5).
\end{equation}
The solution is given by
\begin{equation}
A=\frac{3}{4}\ln (b-3)-\ln (b-1)-\frac{1}{4}\ln (1+b).
\end{equation}
Near the UV point with (6,3) supersymmetries, the fluctuation
behaves as
\begin{equation}
\delta a_1\sim e^{-\frac{r}{2L}},\qquad L=\frac{1}{8g^2}\, .
\end{equation}
The flow is driven by a relevant operator of dimension
$\frac{3}{2}$. At the IR (0,2) point, we find
\begin{equation}
\delta a_1\sim e^{\frac{4r}{3L}},\qquad L=\frac{1}{12g}\, .
\end{equation}
The operator becomes irrelevant with dimension
$\Delta=\frac{10}{3}$. The ratio of the central charges is
\begin{equation}
\frac{c_{\textrm{UV}}}{c_{\textrm{IR}}}=\frac{3}{2}.
\end{equation}
\indent We then move to a flow with two scalars. With
\begin{equation}
a_1=\cosh^{-1}b_1,\qquad a_2=\cosh^{-1}b_2,
\end{equation}
the flow equations are given by
\begin{eqnarray}
b_1'&=&g(b_1^2-1)(b_1-5-b_2+b_1b_2),\\
b_2'&=&\frac{g}{2}(b_2^2-1)[b_1^2(1+b_2)-7+5b_2-2b_1(3+b_2)].
\end{eqnarray}
Taking $b_1$ as a function of $z=b_2$, we find
\begin{equation}
\frac{db_1}{dz}=\frac{2 (-5-z+(1+z) b_1) (-1+b_1^2)}{(-1+z^2) (-7+5
z-2 (3+z) b_1+(1+z) b_1^2)}\,.
\end{equation}
The numerical solution is given in Figure \ref{sol1_36}. The metric
function $A$ can be determined by using the equation
\begin{equation}
\frac{dA}{dz}=-\frac{-3-14 z+5 z^2-2 (5+6 z+z^2) b_1+(1+z)^2
b_1^2}{2 (-1+z^2) (-7+5 z-2 (3+z) b_1+(1+z) b_1^2)}\, .
\end{equation}
The numerical solution is given in Figure \ref{sol2_36}. The
linearized equations give
\begin{equation}
\delta a_1 \sim \delta a_2 \sim e^{-\frac{r}{2L}},\qquad
L=\frac{1}{8g}
\end{equation}
near the UV point. The flow is driven by a relevant operator of
dimension $\frac{3}{2}$ and interpolates between (6,3) and (1,3)
critical points. Near the IR, we find
\begin{equation}
\delta a_1 \sim \delta a_2 \sim e^{\frac{r}{2L}},\qquad
L=\frac{1}{16g}\, .
\end{equation}
So, in the IR the operator has dimension $\frac{5}{2}$. The ratio of
the central charges is
\begin{equation}
\frac{c_{\textrm{UV}}}{c_{\textrm{IR}}}=2.
\end{equation}
\subsubsection{$SO(5)\times SO(4)$ gauging}
In this gauging, there is no critical point with two non zero
scalars, so there is no flow with two scalars. The flow equation
with one scalar is
\begin{equation}
a_1'=g\sinh a_1 (\cosh a_1-5).
\end{equation}
The solution for $r$ as a function of $b=\cosh a_1$ is
\begin{equation}
r=\frac{1}{24g}\ln (b-5)-\frac{1}{8g}\ln (b-1)+\frac{1}{12g}\ln
(1+b).
\end{equation}
The gravitino variation gives
\begin{equation}
A'=-\frac{g}{4}(\cosh (2a_2)-20\cosh a_1-13).
\end{equation}
The solution for $A$ as a function of $b$ is
\begin{equation}
A=\frac{2}{3}\ln (b-5)-\ln (b-1)-\frac{1}{6}\ln [20(1+b)].
\end{equation}
We find the scalar fluctuations near the UV and IR point as
\begin{eqnarray}
\textrm{UV}&:&\,\,\, \delta a_1\sim e^{-\frac{r}{2L}}, \qquad
L_{\textrm{UV}}= \frac{1}{8g},\\
\textrm{IR}&:&\,\,\, \delta a_1\sim e^{\frac{3r}{2L}}, \qquad
L_{\textrm{IR}}= \frac{1}{16g}\, .
\end{eqnarray}
From these, we find that the flow is driven by a relevant operator
of dimension $\frac{3}{2}$. The operator has dimension $\frac{7}{2}$
in the IR. The ratio of the central charges is
\begin{equation}
\frac{c_{\textrm{UV}}}{c_{\textrm{IR}}}=2.
\end{equation}
\subsection{RG flows in non-compact gaugings}
\subsubsection{$G_{2(-14)}\times SL(2)$ gauging}
\begin{figure}[!h] \centering
  \includegraphics[width=0.5\textwidth, bb = 0 0 200 150 ]{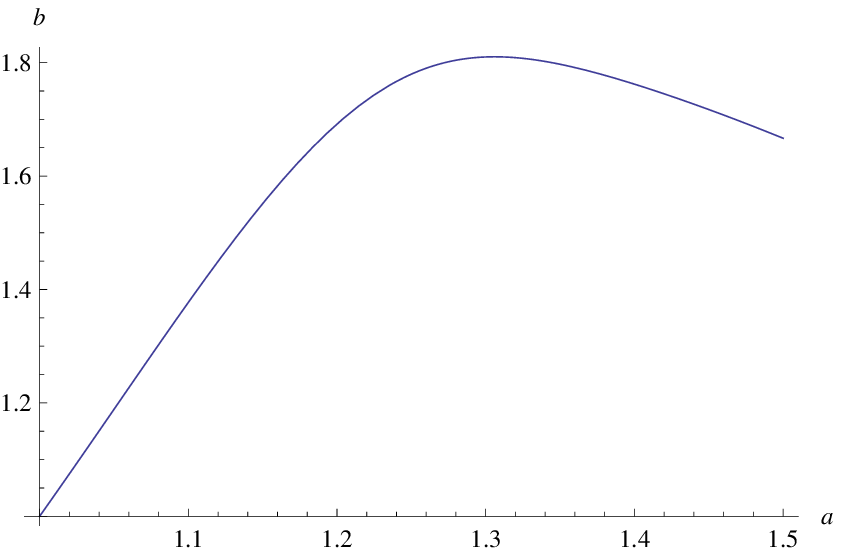}\\
  \caption{Solution for $b(a)$ in $G_{2(-14)}\times SL(2)$ gauging.}
  \label{sol1_G2}
\end{figure}

\begin{figure}[!h] \centering
  \includegraphics[width=0.5\textwidth, bb = 0 0 200 150 ]{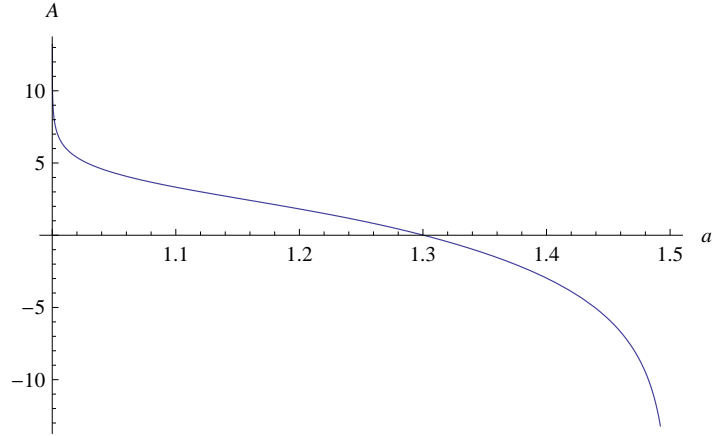}\\
  \caption{Solution for $A(a)$ in $G_{2(-14)}\times SL(2)$ gauging.}
  \label{sol2_G2}
\end{figure}

Remarkably, there exist flow solutions in this non-compact
exceptional gauging. We start with a single scalar giving rise to
the flow equation
\begin{equation}
a_1'=-\frac{4g}{3}\sinh a_1(\cosh a_1-2).
\end{equation}
The solution to this equation is
\begin{equation}
r=\frac{3}{4g}\big[\frac{1}{3}\ln (b-2)-\frac{1}{2}\ln
(b-1)+\frac{1}{6}\ln (1+b)\big]
\end{equation}
where as usual $b=\cosh a_1$. The equation for $A$ and its solution
are given by
\begin{eqnarray}
A'&=&\frac{8g}{3}\cosh a_1-\frac{2g}{3}\sinh^2 a_1 \\
\textrm{and}\,\,\, A&=&\frac{5\ln (b-2)-6\ln (b-1)-2\ln(1+b)}{6}\, .
\end{eqnarray}
The solution interpolates between (7,2) and (0,1) critical points
with the ratio of the central charges
\begin{equation}
\frac{c_{\textrm{UV}}}{c_{\textrm{IR}}}=\frac{5}{4}\, .
\end{equation}
The linearized equation gives
\begin{eqnarray}
\textrm{UV}&:&\,\,\, \delta a_1\sim e^{-\frac{r}{2L}},\qquad
L_{\textrm{UV}}=\frac{3}{8g},\\
\textrm{IR}&:&\,\,\, \delta a_1\sim e^{\frac{78r}{5L}},\qquad
L_{\textrm{IR}}=\frac{3}{10g}\, .
\end{eqnarray}
The flow is driven by a relevant operator of dimension
$\frac{3}{2}$. In the IR, the operator has dimension $\frac{88}{5}$.
\\ \indent We now move to a flow solution with two scalars. The flow
equations are
\begin{eqnarray}
a_1'&=&-\frac{2g}{3}\big[4\sinh
a_2-\cosh^2\frac{a_2}{3}\sinh(2a_1)\big],\\
a_2'&=&-\frac{g}{12}[2(9+8\cosh a_1-\cosh (2a_1))\sinh a_2\nonumber
\\ & &-(7+\cosh (2a_1))\sinh{(2a_2)}].
\end{eqnarray}
Using
\begin{equation}
a_1=\cosh^{-1}a,\qquad a_2=\cosh^{-1} b,
\end{equation}
we can combine the two equations into one with $a$ being independent
variable
\begin{equation}
\frac{db}{da}=\frac{(b^2-1)(b(3+a^2)+a^2-4a-5)}{2(a^2-1)(ab+a-4)}\,
.
\end{equation}
We give a numerical solution to this equation in Figure
\ref{sol1_G2}. The equation for $A$ is
\begin{equation}
\frac{dA}{da}=-\frac{3-8a+a^2+2(a^2-4a-5)b+(3+a^2)b^2}{4(a^2-1)(ab+a-4)}
\end{equation}
whose solution is shown in Figure \ref{sol2_G2}. The scalar
fluctuations are given by
\begin{eqnarray}
\textrm{UV}&:&\,\,\, \delta a_1\sim\delta a_2\sim e^{\frac{-r}{2L}},
\qquad
L_{\textrm{UV}}= \frac{3}{8g},\\
\textrm{IR}&:&\,\,\, \delta a_1\sim\delta a_2\sim e^{\frac{r}{4L}},
\qquad L_{\textrm{IR}}= \frac{9}{32g}\, .
\end{eqnarray}
The flow interpolates between the (7,2) and (1,2) points with the
ratio of the central charges
\begin{equation}
\frac{c_{\textrm{UV}}}{c_{\textrm{IR}}}=\frac{4}{3}\, .
\end{equation}
The flow is driven by a relevant operator of dimension
$\frac{3}{2}$.
\section{Conclusions}\label{conclusion}
We have studied $N=9$ three dimensional gauged supergravity with
compact and non-compact gaugings. We have found some supersymmetric
AdS$_3$ vacua corresponding to some two dimensional CFT's. We have
identified the superconformal groups from the isometry of the
AdS$_3$ backgrounds with $L=\mathbf{I}$. These backgrounds have dual
conformal field theories at their boundaries. We then studied RG
flow solutions describing a deformation of the CFT in the UV to the
CFT in the IR. In the scalar sector studied here, only $SO(7)\times
SO(2)$, $SO(6)\times SO(3)$, $SO(5)\times SO(4)$ and
$G_{2(-14)}\times SL(2)$ gaugings admit supersymmetric flow
solutions. This is because there is only one supersymmetric critical
point in $SO(9)$, $SO(8)$ and $Sp(1,2)\times SU(2)$ gaugings. This
is not unexpected, the bigger gauge groups give rise to a simpler
structure of vacua in general. We have found analytic flow solutions
with one active scalar and numerical solutions for the flows with
two active scalars. All the flows are operator flows driven by a
relevant operator of dimension $\frac{3}{2}$. It is interesting to
identify the CFT's dual to these gravity solutions. Because two
dimensional field theories are more controllable and the gravity
solutions correspond to strong coupling limits of the dual field
theories, we hope to understand many aspects of the AdS/CFT
correspondence in the case of AdS$_3$/CFT$_2$.
\\ \indent
The higher dimensional origin of many three dimensional gauged
supergravities is still mysterious. Only the case of non semisimple
gaugings is known to be related to dimensional reductions of higher
dimensional theories \cite{csym}. It is interesting to study the non
semisimple gaugings in this $N=9$ theory although there is another
subtlety with the theories with odd $N$. This is because we cannot
obtain these theories directly from dimensional reductions due to
the mismatch in the number of supercharges. The reduced theory,
always having even $N$ in three dimensions, needs to be truncated in
order to give odd values of $N$. The models with compact and
non-compact gauge groups studied in this paper and elsewhere are not
obtainable from dimensional reductions, so it is very interesting to
study whether there exist any higher dimensional origin for these
models. This will provide an interpretation of our flow solutions
and that studied in \cite{bs} in terms of higher dimensional
geometries.
\acknowledgments We would like to thank Henning Samtleben and K. S.
Narain for valuable discussions and Ahpisit Ungkitchanukit for
reading the manuscript.
\appendix
\section{Essential formulae}\label{detail}
In this appendix, we give all necessary formulae in order to obtain
the scalar potential and flow equations. We use the 52 generators of
$F_4$ from \cite{F4}. The generators are normalized by
\begin{equation}
\textrm{Tr}(c_ic_j)=-6\delta_{ij}.
\end{equation}
With this normalization, we find that
\begin{eqnarray}
\mathcal{V}^{\alpha IJ}&=&-\frac{1}{6}\textrm{Tr}(L^{-1}T_G^\alpha
LX^{IJ}) \\
\mathcal{V}^{\alpha A}&=&\frac{1}{6}\textrm{Tr}(L^{-1}T_G^\alpha
LY^{A})\label{Vcoset}
\end{eqnarray}
where we have introduced the symbol $T^\alpha_G$ for gauge group
generators. $T^\alpha_G$ will be replaced by some appropriate
generators of the gauge group being considered in each gauging.
\\ \indent The following mapping provides the relation between $c_i$ and
$X^{IJ}$, generators of $SO(9)$,
\begin{eqnarray}
X^{1 2}&=& c_1,\,\, \, X^{1 3}= -c_2,\,\, \, X^{2 3}= c_3,\,\, \,
X^{3 4}=
c_6,\,\, \, X^{1 4}= c_4,\,\, \, X^{2 4}= -c_5,\nonumber \\
X^{1 5}&=& c_7,\,\, \, X^{2 5}= -c_8,\,\,\, X^{3 5}= c_9,\,\, \,
X^{4 5}= -c_{10},\,\, \, X^{5 6}= -c_{15},\,\, \, X^{1 6}= c_{11},
\nonumber \\
 X^{2  6}&=&
-c_{12},\,\, \, X^{4 6}= -c_{14},\,\, \, X^{3 6}= c_{13},\,\,\, X^{1
7}= c_{16},\,\, \, X^{27}= -c_{17},\,\, \, X^{4 7}= -c_{19}, \nonumber \\
X^{3 7}&=& c_{18},\,\, \, X^{6 7}= -c_{21},\,\, \, X^{ 5 7}=
-c_{20},\,\, \, X^{7 8}= -c_{36},\,\,\, X^{1 8}= c_{30},\,\, \, X^{2
8}= -c_{31},
\nonumber \\
X^{4 8}&=& -c_{33},\,\, \, X^{3 8}= c_{32},\,\, \, X^{6 8}=
-c_{35},\,\, \, X^{5 8}= -c_{34},\,\, \, X^{2 9}= -c_{46},\,\,\,
X^{1 9}= c_{45},
\nonumber \\
X^{4 9}&=& -c_{48},\,\,\, X^{3 9}= c_{47},\,\,\, X^{6 9}=
-c_{50},\,\, \, X^{5 9}= -c_{49},\nonumber \\ X^{8 9}&=&
-c_{52},\,\, \, X^{7 9}= -c_{51}\, .
\end{eqnarray}
All the $f^{IJ}$'s components can be obtained from the structure
constants of the $[X^{IJ},Y^A]$ given in \cite{F4}, but we will not
repeat them here. \\ \indent In the non-compact $G_{2(-14)}\times
SL(2)$ gauging, we use the following generators. The generators of
$G_{2(-14)}$ are obtained by using the embedding of $G_{2(-14)}$ in
$SO(7)$ generated by $X^{IJ}$, $I,J =1,\ldots , 7$. The adjoint
representation of $SO(7)$ decomposes under $G_{2(-14)}$ as
\begin{equation}
\mathbf{21}\rightarrow \mathbf{14}+\mathbf{7}\, .
\end{equation}
The generators of $G_{2(-14)}$ can be explicitly found by
combinations of $SO(7)$ generators \cite{G2inSO(7)}
\begin{eqnarray}
T_1&=&\frac{1}{\sqrt{2}}(X^{36}+X^{41}),\,\,\,T_2=\frac{1}{\sqrt{2}}(X^{31}-X^{46}),\nonumber
\\
T_3&=&\frac{1}{\sqrt{2}}(X^{43}-X^{16}),\,\,\,T_4=\frac{1}{\sqrt{2}}(X^{73}-X^{24}),\nonumber
\\
T_5&=&-\frac{1}{\sqrt{2}}(X^{23}+X^{47}),\,\,\,T_6=-\frac{1}{\sqrt{2}}(X^{26}+X^{71}),\nonumber
\\T_7&=&\frac{1}{\sqrt{2}}(X^{76}-X^{21}),\,\,\, T_8=\frac{1}{\sqrt{6}}(X^{16}+X^{43}-2X^{72}),\nonumber
\\
T_9&=&-\frac{1}{\sqrt{6}}(X^{41}-X^{36}+2X^{25}),\,\,\,
T_{10}=-\frac{1}{\sqrt{6}}(X^{31}+X^{46}-2X^{57}),\nonumber
\\T_{11}&=&\frac{1}{\sqrt{6}}(X^{73}+X^{24}+2X^{15}),\,\,\,
T_{12}=-\frac{1}{\sqrt{6}}(X^{74}-X^{23}+2X^{65}),\nonumber
\\T_{13}&=&\frac{1}{\sqrt{6}}(X^{26}-X^{71}+2X^{35}),\,\,\,
T_{14}=\frac{1}{\sqrt{6}}(X^{21}+X^{76}-2X^{45}).
\end{eqnarray}
We have verified that these generators satisfy $G_2$ algebra given
in \cite{G2}. The $SL(2)$ generators are
\begin{equation}
J_1=i\sqrt{2}(c_{22}+c_{27}),\,\,\,
J_2=i\sqrt{2}(c_{37}+c_{42}),\,\,\, J_3=2c_{52}
\end{equation}
which can be easily checked that they commute with all $T$'s and
form $SL(2)$ algebra. \\ \indent The generators of non-compact
$Sp(1,2)$ can be constructed by first finding its compact subgroup
generators $Sp(1)\times Sp(2)\sim SO(3)\times SO(5)$. The latter can
be obtained by taking $SO(8)$ with generators $X^{IJ}$, $I,
J=1,\ldots, 8 $. We then identify the $SO(3)$ generators with
$X^{IJ}$ for $I, J=1,\ldots, 3 $ and $SO(5)$ with $X^{IJ}$ for $I,
J=4,\ldots, 8 $. The eight non-compact generators of $Sp(1,2)$ can
be obtained by taking combinations of $Y^A$'s which commute with the
$SU(2)$ gauge group. The latter has three generators obtained by
looking for the combinations of $SO(9)$ generators that commute with
$SO(3)\times SO(5)$ mentioned above. We find the following gauge
generators:
\begin{itemize}
  \item Sp(1,2):
  \begin{eqnarray}
  Q_1&=&\sqrt{2}c_1,\,\,\,Q_2=-\sqrt{2}c_2,\,\,\,Q_3=\sqrt{2}c_3,\,\,\,Q_4=\sqrt{2}c_4,\,\,\,Q_5=-\sqrt{2}c_5,\nonumber
  \\Q_6&=&\sqrt{2}c_6,\,\,\,Q_7=\sqrt{2}c_7,\,\,\,Q_8=-\sqrt{2}c_8,\,\,\,Q_9=\sqrt{2}c_9,\,\,\,Q_{10}=-\sqrt{2}c_{10},\,\,\,\nonumber
  \\Q_{11}&=&-c_{21}-c_{52},\,\,\,Q_{12}=c_{51}-c_{35},\,\,\,Q_{13}=c_{50}+c_{36},\nonumber
  \\Q_{14}&=&Y_1+Y_{10},\,\,\,Q_{15}=Y_2-Y_{9},\,\,\,Q_{16}=Y_3+Y_{13},\,\,\,
  \nonumber \\
  Q_{17}&=&Y_4+Y_{16},\,\,\,Q_{18}=Y_5-Y_{11},\,\,\,Q_{19}=Y_6-Y_{15},\nonumber
  \\
  Q_{20}&=&Y_7+Y_{14},\,\,\,Q_{21}=Y_8-Y_{12}\, .
  \end{eqnarray}
  \item SU(2):
  \begin{equation}
  K_1=\frac{1}{2}(c_{52}-c_{21}),\,\,\,
  K_2=-\frac{1}{2}(c_{35}+c_{51}),\,\,\,
  K_3=\frac{1}{2}(c_{36}-c_{50}).
  \end{equation}
\end{itemize}
With these generators and \eqref{Vcoset}, we can compute the
T-tensors
\begin{eqnarray}
T^{IJ,KL}&=&\mathcal{V}^{IJ,\alpha} \mathcal{V}^{KL,\beta}
\delta^{SO(p)}_{\alpha\beta}
-\mathcal{V}^{IJ,\alpha} \mathcal{V}^{KL,\beta} \delta^{SO(9-p)}_{\alpha\beta}, \\
T^{IJ,A}&=&\mathcal{V}^{IJ,\alpha} \mathcal{V}^{A,\beta}
\delta^{SO(p)}_{\alpha\beta} -\mathcal{V}^{IJ,\alpha}
\mathcal{V}^{A,\beta} \delta^{SO(9-p)}_{\alpha\beta}
\end{eqnarray}
for compact gaugings and
\begin{eqnarray}
T^{IJ,KL}&=&\mathcal{V}^{IJ,\alpha} \mathcal{V}^{KL,\beta}
\eta^{G_1}_{\alpha\beta}
-K\mathcal{V}^{IJ,\alpha} \mathcal{V}^{KL,\beta} \eta^{G_2}_{\alpha\beta}, \\
T^{IJ,A}&=&\mathcal{V}^{IJ,\alpha} \mathcal{V}^{A,\beta}
\eta^{G_1}_{\alpha\beta} -K\mathcal{V}^{IJ,\alpha}
\mathcal{V}^{A,\beta} \eta^{G_2}_{\alpha\beta}
\end{eqnarray}
for non-compact gaugings with $K$ being $\frac{1}{6}$ and $12$ for
$G_1\times G_2=G_{2(-14)}\times SL(2)$ and $Sp(1,2)\times SU(2)$,
respectively. We have used summation convention over gauge indices
$\alpha$, $\beta$ with the notation $\delta^{G_0}$ and $\eta^{G_0}$
meaning that the summation is restricted to the $G_0$ generators.

\end{document}